\documentclass[english]{cccconf}
\usepackage[comma,numbers,square,sort&compress]{natbib}
\usepackage{epstopdf}
\usepackage{url}

\begin{document}

\title{An Evaluation System for DeFi Lending Protocols}

\author{Shuai Yang\aref{scut},
        Wei Cui\aref{scut,pazhou}}

\affiliation[scut]{The School of Automation Science and
	Engineering, South China University of Technology, Guangzhou 510641, China 
	}
\affiliation[pazhou]{Pazhou Laboratory, Guangzhou, 510330, China \email{aucuiwei@scut.edu.cn}}

\maketitle

\begin{abstract}
With the development of decentralized finance (DeFi), lending protocols have been increasingly proposed in the market.
A comprehensive and in-depth evaluation of lending protocol
is essential to the DeFi market participants. Due to the
short development time of DeFi, the current evaluation is limited
to the single evaluation indicator, such as total locked value (TVL). To our knowledge, we are
the first to study the evaluation model for DeFi lending protocol.
In this paper, we build a four-layer lending protocol evaluation model
based on Analytic Hierarchy Process (AHP). The model contains
three first-level indicators and ten second-level indicators,
covering various aspects of the performance of lending protocol.
We calculated these indicators by obtaining on-chain and off-chain data of lending protocol. Then we evaluated and
ranked six mainstream lending protocols utilizing the proposed
model during the period 2021/8/20-2022/11/10. Through comparative analysis of evaluation result, we found that the
violent decline in the price of ETH increase the market share of stablecoin in the lending protocols. In addition, we also revealed the reasons for various fluctuations. 
\end{abstract}


\footnotetext{This work was supported by the National Key Research and Development Program of China under Grant 2022YFB3103100, and the National Natural Science Foundation of China under Grant 62273154. Corresponding author: Wei Cui.}

\section{Introduction}

Financial architecture based on blockchain and distributed ledger technology, so called decentralized finance (DeFi), has developed rapidly from June 2020. Compare to centralized finance, the advantage of DeFi includes permissionless, trustless,  transparent and interconnectedness \cite{cefi,santos,Salami} . The financial services provided by DeFi can be divided into five categories: decentralized exchanges (DEXes), lending platforms, asset handling platforms, derivative services and payment networks \cite{DeFi-ning DeFi?Challenges & Pathway,SoK?Decentralized Finance (DeFi),Busayatananphon,Schar}. Lending is a very important component in DeFi. As of May 2022, the total locked value of lending platforms has reached \$48.63B, accounting for 34.92\% of the total value locked of DeFi \cite{llama}.

DeFi lending protocols allow borrowers to deposit one token as collateral and lend another token for overcollateralization and supplier to supply liquidity for lending pool\cite{loanable funds}. On the one hand, Oracle can provide price information of off-chain DeFi assets\cite{Zhao} \cite{Caldarelli}. On the other hand, Liquidation helps protocols reduce exposure to debt when collateral prices fall \cite{Liquidation}. With the development of DeFi, more and more lending protocols has appeared in the lending market. For the participants of lending protocol, the comprehensive ranking of protocols is very meaningful, and it can help users analysis different protocols quickly and intuitively. 
To our knowledge,  the current method for ranking lending protocols is the single indicator, such as total value locked, market capitalization, etc. However, the information reflected by the single indicator is very limited, and the lending protocols needs to be analyzed from more perspectives. Existing literature analyzes lending protocols from the perspective of the decentralization and market share.

The decentralization of DeFi protocols is a topic worth investigating. When the DeFi protocol is initially running, ERC20-standards tokens will be launched by initialing coin offerings (ICOs) at the same time\cite{Fenu}. In general, holders of token can govern the protocol through governance voting, such as changing the key parameters of the protocol. The distribution of governance tokens also determines the decentralization and safety of the protocol. An attack on the protocol can be launched by holding more than 51\% of the tokens \cite{The decentralized crises}. In order to more conveniently measure the decentralization of cryptocurrency distribution, Gochhayat et al.\cite{d5} proposes five metrics to measure the Decentrality of Bitcoin's governance layer.  Nadler\cite{d3} and Lin\cite{d4} then used these metrics to measure the decentralization of the mainstream protocols in DeFi, and the results showed a very low degree of decentralization. Zhang\cite{d1} summarizes the existing measures of decentralization.

The token assets of DeFi protocol, which can reflect the current market share of the protocol, has also attracted a lot of attention in the past few years. DeFi token assets possess stronger and stable correlations with Ethereum, and DeFi markets are viewed as a separate asset class from conventional cryptocurrencies, such as Bitcoin and Ethereum\cite{Goodell} \cite{Corbet}. In addition, there are certain bubbles in DeFi assets, and some indicators can be used to evaluate whether the current assets are reasonable, such as the total locked value, which is negatively correlated with the bubble of DeFi assets\cite{Maouchi} \cite{Karim} \cite{Chousa}. The suitable evaluation indicator will help to overcome the risk of DeFi token asssets' bubbles. 


In general, lending protocol is a very important component of the DeFi architecture, and protocol participants tend to pay attention to the protocol's market share, valuation, and decentralization. Therefore, the evaluation of the various aspects is the basis for the analysis of lending protocols. By formulating a comprehensive lending protocol evaluation system, the evaluation results of mainstream protocols can be efficiently obtained. Through the analysis of evaluation results, we can discover the deep-seated reasons and trends
of fluctuations in the aspects of lending market, which provide the participants with judgements quickly.

A comprehensive evaluation system is very helpful for analyzing the current status of development and evolution trend of this field. However, there are few studies related to evaluation in the blockchain field \cite{topsis} \cite{Ozili} \cite{Hartmann}.To our knowledge, we are the first to study the evaluation model for DeFi lending protocols. The challenge of evaluating lending protocols lies in how to evaluate protocol from multiple perspectives.
In this paper, we aim to evaluate mainstream lending protocols. Firstly, we build a four-layer lending protocol evaluation model based on Analytic Hierarchy Process (AHP). Secondly, By collect the data of the lending protocols and simulate based on the above evaluation model, we obtain the evaluation results of mainstream lending protocols.

This paper is organized as follows. In section \ref{sec:model}, we establish our evaluation model. Section \ref{sec:data} presents the detail of data colleciton and processing. In section \ref{sec:result}, we analyze and discuss the experiment results. Section \ref{sec:conclusion} concludes this paper.

\section{DeFi Ranking Model}\label{sec:model}
\begin{figure}[h]
	\centering
	\includegraphics[width=13cm,trim=70 20 50 10]{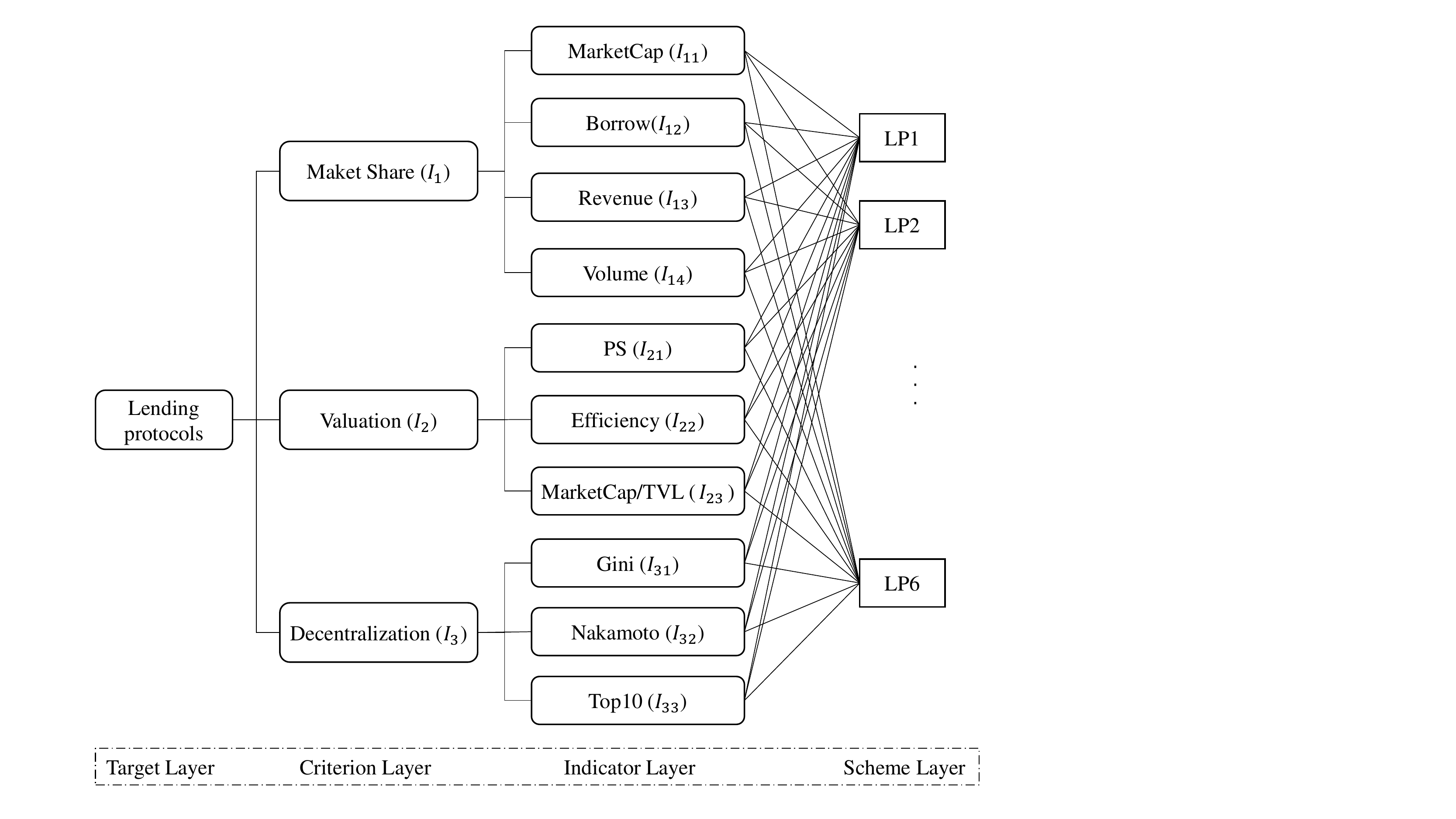}
	\caption{Ranking model architecture for lending protocol.}
	\label{fig:model}
\end{figure}

As shown in figure \ref{fig:model}, we established a four-layer ranking model based on AHP. The highest level is target layer and the target is to select the best DeFi lending protocol. The middle layer is the criterion layer and indicator layer, which include all indicators for evaluating DeFi lending protocols. The bottom layer is the scheme layer, including all DeFi lending protocols, and the result of the decision will be selected from the scheme layer. In the criterion layer, we use three first-level indicators and ten second-level indicators to evaluate the protocol in different aspects. The specific meaning of indicators are as follows:

${I_1}$:Market Share

${I_{11}}$:Market Capitalization

${I_{12}}$:Total borrow amount of protocol

${I_{13}}$:Daily revenue of protocol

${I_{14}}$:Daily volume of token of protocol

${I_2}$:Protocol Valuation

${I_{21}}$:The price-to-sales ratio of protocol

${I_{22}}$:The capital efficiency of protocol

${I_{23}}$:The Market Capitalization to total value locked ratio of protocol

${I_3}$:Decentralized

${I_{31}}$:The Gini coefficient of token distribution

${I_{32}}$:The Nakamoto coefficient of token distribution

${I_{33}}$:The proportion of tokens held by top ten holders

The indicators are the core of the evaluation model. The first principle of our indicator selection is to have data basis, and then to be comprehensive and reasonable. Our proposed evaluation model is effective as long as the indicator data is available. We choose first-level indicators mainly considering the regular lending protocols market share, protocol valuation and decentralization. The second-level indicators explain the first-level indicators in detail.

\subsection{Ranking indicators}
\textit{ 1) Market Share(${I_1}$)}: The market share indicator reflects the share of protocol in the field of lending. Market share includes four secondary indicators. Indicator ${I_{11}}$ is the market capitalization of protocol, which is equal to the product of the current price of the token and the amount of tokens currently in circulation. 
Indicator ${I_{12}}$ and Indicator ${I_{13}}$ respectively represent the total borrow amount and the daily revenue of the lending protocol. The borrow amount is a very important indicator for the lending protocol. On the one hand, the revenue of the protocol comes from the interest paid by the borrower when repaying the debt. On the other hand, based on the incentive of the interest rate model of the lending protocol, the increase of borrow amount can attract more and more borowers and depositors and locked more assets. Finally, borrow amount and deposit amount reach a dynamic equilibrium point. In addition, although the greater the number of protocol borrow, the greater the protocol revenue, due to the different interest rate parameters of each lending pool and the different time for borrowers to repay their debts, there is no correlation between the indicator of borrow and revenue. We regard that the above four secondary indicators are positively correlated with the first indicator of market share.
Indicator ${I_{14}}$ represents the daily trading volume of the protocol token, which can reflect the liquidity of the token in the market. The higher the token trading volume, the better the market liquidity of the protocol token.

\textit{ 2) Valuation(${I_2}$)}:
The valuation indicator can reflect whether the token price of protocol is reasonable. There are three secondary indicators under this primary indicator.
Indicator ${I_{21}}$ refer to the price-to-sale ratio(ps), which is an indicator for evaluating the value of stocks in traditional finance, and is also applicable to the valuation of DeFi protocols. Price-to-sales ratio is equal to the fully diluted market cap divided by annualized total revenue. Indicator ${I_{22}}$ refers to the capital efficiency of the protocol, which can reflect the revenue obtained by the protocol through the total value locked. The indicator of efficiency is equal to the ratio of the protocol total value locked to the protocol revenue. Indicator ${I_{23}}$ is the market capitalization to total value locked ratio. Generally, a lending protocol with a high total value locked will have a higher market capitalization. When the indicator is less than or equal to 1, it means that the protocol may be underestimated. We regard both indicator ${I_{21}}$ and ${I_{23}}$ as a positive correlation with valuation and ${I_{22}}$ as a negative correlation.

\textit{ 3) Decentralization(${I_3}$)}: The indicator of decentralization measures the distribution of goverance token of lending protocol. The more centralized the token distribution, the less decentralized the protocol. It includes three secondary indicators. Indicator ${I_{31}}$ is Gini coefficent of token distribution. Gini coefficent can reflect the overall distribution of protocol tokens. The indicator ${I_{32}}$ refer to Nakamoto coefficient that the minimum number of holders who hold more than 50\% of the tokens. We regard indicators ${I_{32}}$ as a positive correlation with decentralization. Arrange addresses in descending order according to the number of tokens held, and indicator ${I_{33}}$ is the proportion of tokens held by the top ten addresses. Both indicator ${I_{31}}$ and ${I_{33}}$ are negative correlation with decentralization, so it is processed by subtracting the value of indicator from 1. 

\subsection{Ranking process}
In this section, we will introduce the specific evaluation process of the AHP model. 

\textit{Step 1: Construct a set of pair-wise comparison matrices}. A total of 10 matrices need to be constructed according to the 9-level importance scale of AHP. Among them, matrices involved in scheme layer
were constructed based on the indicator scores of lending protocol. The rest of pair compare matrices were constructed
according to user-defined weights. 

\textit{Step 2: Consistency Check}. Having made all the pair-wise comparisons, the
consistency is determined by using the eigenvalue,${\lambda _{max}}$, to calculate the consistency index.
\begin{equation}
	CI = \frac{{{\lambda _{max}} - n}}{{n - 1}},
\end{equation}
where $n$ is the matrix size.CI is used to measure the degree of consistency of pair
compare matrix. RI denotes the average random consistency
index. Next we calculated the consistency ratio:
\begin{equation}
	CR = \frac{{CI}}{{RI}} < \frac{1}{10}.
\end{equation}	
CR is the ratio of CI to RI , which is called the random
consistency ratio. If CR is less than 0.1, the pair compair
matrix pass the consistency test, otherwise it needs to be
adjusted.

\textit{Step 3: Score}. After finishing the consistency check, we
obtained the weight vector by normalizing the eigenvector
corresponding to the largest eigenvalue. ${w_{ij}}$ denotes the weight
of the j-th second-level indicator of the ${{l_i}}$ first-level indicator
to the i-th first-level indicator. Let ${{l_i}}$ denote the number of
second-level indicators of the i-th first-level indicator. ${x_{ij}}$
denotes the score of the j-th second-level indicator under the
i-th first-level indicator of lending protocol. Then we got lending protocol's score of the i-th first-level indicator:
\begin{equation}
	{c_i} = \sum\limits_{j = 1}^{{l_i}} {{w_{ij}} \cdot {x_{ij}}} 
\end{equation}
where ${w_i}$ denotes the weight of the i-th first-level indicator to the
target, and ${c_i}$ denotes the score of the i-th first-level indicator
of lending protocol. So we got lending protocol's final score by the
following formula:
\begin{equation}
	score = \sum\limits_{i = 1}^4 {{w_i} \cdot {c_i}} 
\end{equation}

\section{Data collection and process}\label{sec:data}
In this section, We first introduce the lending protocol for ranking.Then we describe our data collection methodology. Based on these data, we complete the construction of the indicator.

As shown in table\ref{tab:lending}, we select top six lending protocols based on total value locked in DeFiLlama. We choose the above protocols for the following reasons: First, compared with some newly released lending protocols, the above six protocols have been launched for a long time and have sufficient data, which is a key factor in analyzing changes in protocol rankings. Secondly, the number of users and total locked value of these protocols are higher than other protocols, which is more valuable and meaningful for analysis.

\begin{table}[!htbp]
	\centering
	\label{tab:lending}
	\caption{Lending protocols}
	\begin{tabular}{c|c|c}
		\hhline
		&Token&Genesis Date\\ \hline
		
		Aave&AAVE&2020-10-02\\ \hline
		
		Compound&COMP&2020-06-14\\ \hline
		
		MakerDAO&MKR&2017-12-15\\ \hline
		
		Alpha Homora&ALPHA&2020-09-27\\ \hline
		
		Liquity&LQTY&2021-04-05\\ \hline
		
		Benqi&QI&2021-08-11\\
		\hhline
	\end{tabular}
\end{table}

We obtained the on-chain data of protocol tokens through Etherscan\cite{Etherscan} from the genesis date to December 2022. Based on token transfer data collected, we can obtain token distribution at any time to calculate the scores of the decentralization-related indicators. In addition, we found that exchange addresses, such as coinbase, Binance and Huobi, and contract addresses hold lots of tokens. However, these addresses do not vote on community governance and have not an impact on the decentralization of the protocol, and then we filter out both the contract address and the exchange address. Moreover, we excluded addresses that hold token amount less than 10 USD. Given the fact that execute vote require the transaction fees, these addresses vary rarely participate in governance processes. In addition, due to the lack of token price on the chain, we obtained relevant metrics data through Tokenterminal.com\cite{tokenterminal}, such as market capitalization, borrow amount, etc. Based on the off-chain data, we can construct the market share and valuation related indicators.

\section{Result}\label{sec:result}
Based on the ranking model proposed in section \ref{sec:data}, we calculated the scores of six lending protocols from August 2021 to November 2022. To remain objective, all custom weights are initialized to 1. In this section, we first discussed the evolution of protocol scores. Then, by comparing the evolution of first-level indicators with lending protocols, we explained the advantages of evaluation model. Finally, through comparative analysis of second-level indicators, we found lots of in-depth information about lending protocols.
\begin{figure}[ht]
	\centering
	\includegraphics[width=8.7cm,trim=30 0 0 0]	{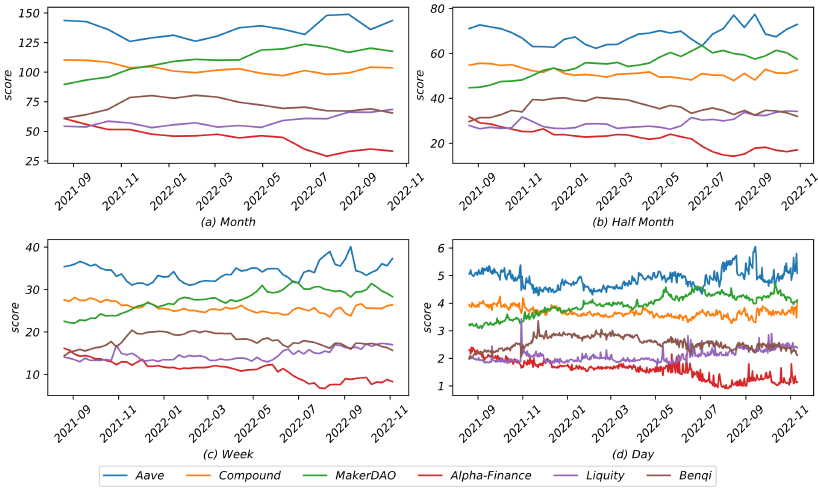}
	\caption{The evoluation of lending protocol scores under different time granularities.}
	\label{fig:score}
\end{figure}

\begin{figure}[t]
	\centering
	\includegraphics[width=9cm,trim=50 0 0 0]{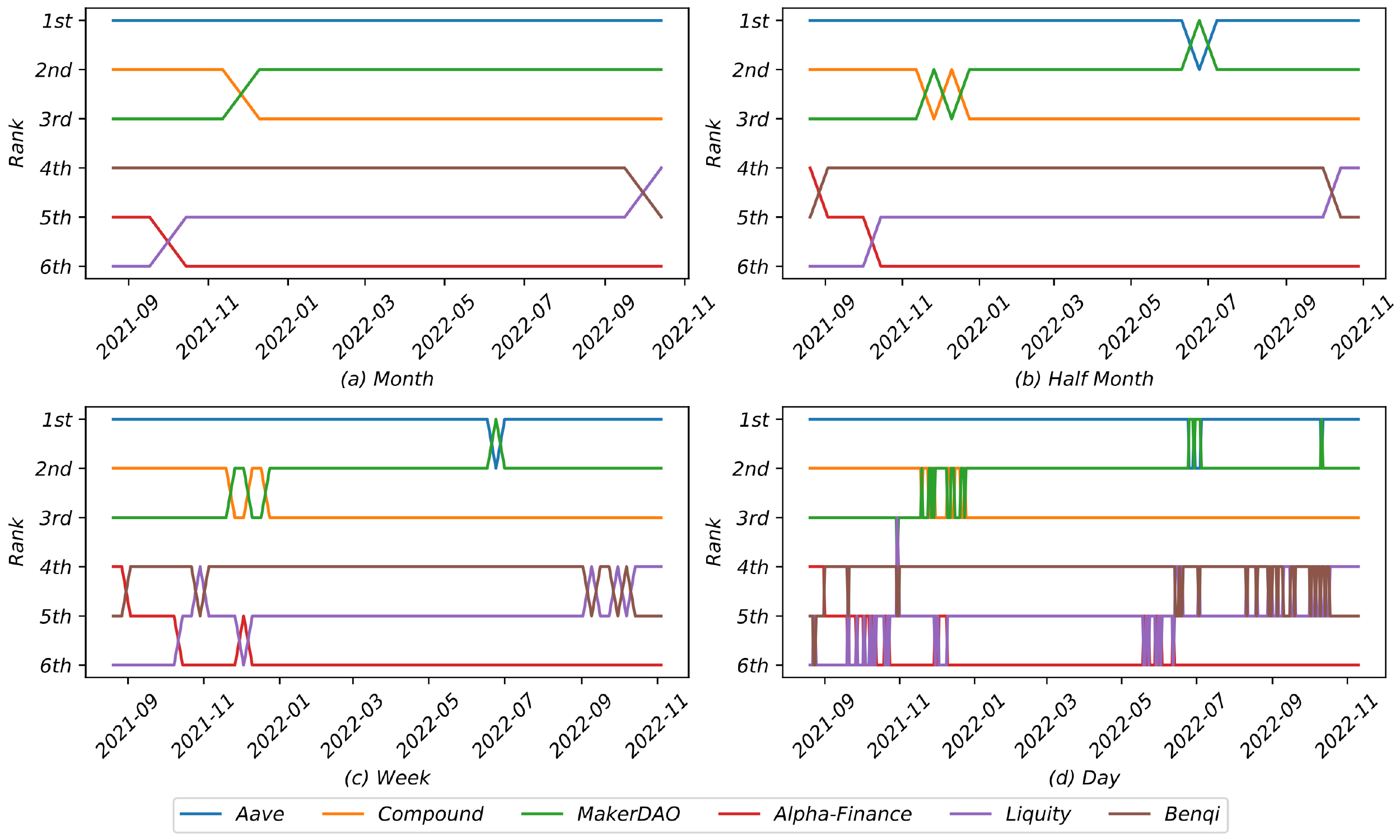}
	\caption{The evoluation of lending protocol rankings under different time granularities.}
	\label{fig:rank}
\end{figure}
\subsection{Selection of Time Granularity and Ordinate}
\textit{ 1) Time Granularity:} Comparison of lending protocol scores evolution under different time granularities is shown in \ref{fig:score}. Generally speaking, the finer the time granularity of graph is, the more information it contains, and the more difficult it is to analyze. For example, it is difficult to quickly figure out the evolution trend from the figure \ref{fig:score}(d). Conversely, the coarsening of time granularity is accompanied by the loss of information. For instance, lots of short-term fluctuations in figure \ref{fig:score}(b) and figure \ref{fig:score}(c) are smoothed out in figure
\ref{fig:score}(a). In the balance between accuracy and difficulty, half a month or week is a good choice. Because week and half a month have similar analysis difficultysince and week contains more information, we select week as the time granularity of subsequent analysis.

\textit{ 2) Ordinate:} Compared to figure \ref{fig:score}, figure \ref{fig:rank} uses ranking as the ordinate. By limiting fluctuations to specific rankings, figure 
\ref{fig:rank} can reflect the impact of fluctuations more intuitively. In addition, ranking can clearly show the division and composition of each echelon of lending protocol. For example, figure \ref{fig:rank}(a) clearly shows that Aave is in the first place most of time, MakerDao and Compound form the second echelon.
Therefore, ranking is suitable as the ordinate of macro analysis. So we use ranking as the evaluation result in our software. Score is suitable as the ordinate of micro-analysis and can reveal more information. It is choosed as the ordinate of graphs in the subsequent analysis.

\begin{figure}[h]
	\centering
	\includegraphics[width=9cm,trim=30 0 0 0]{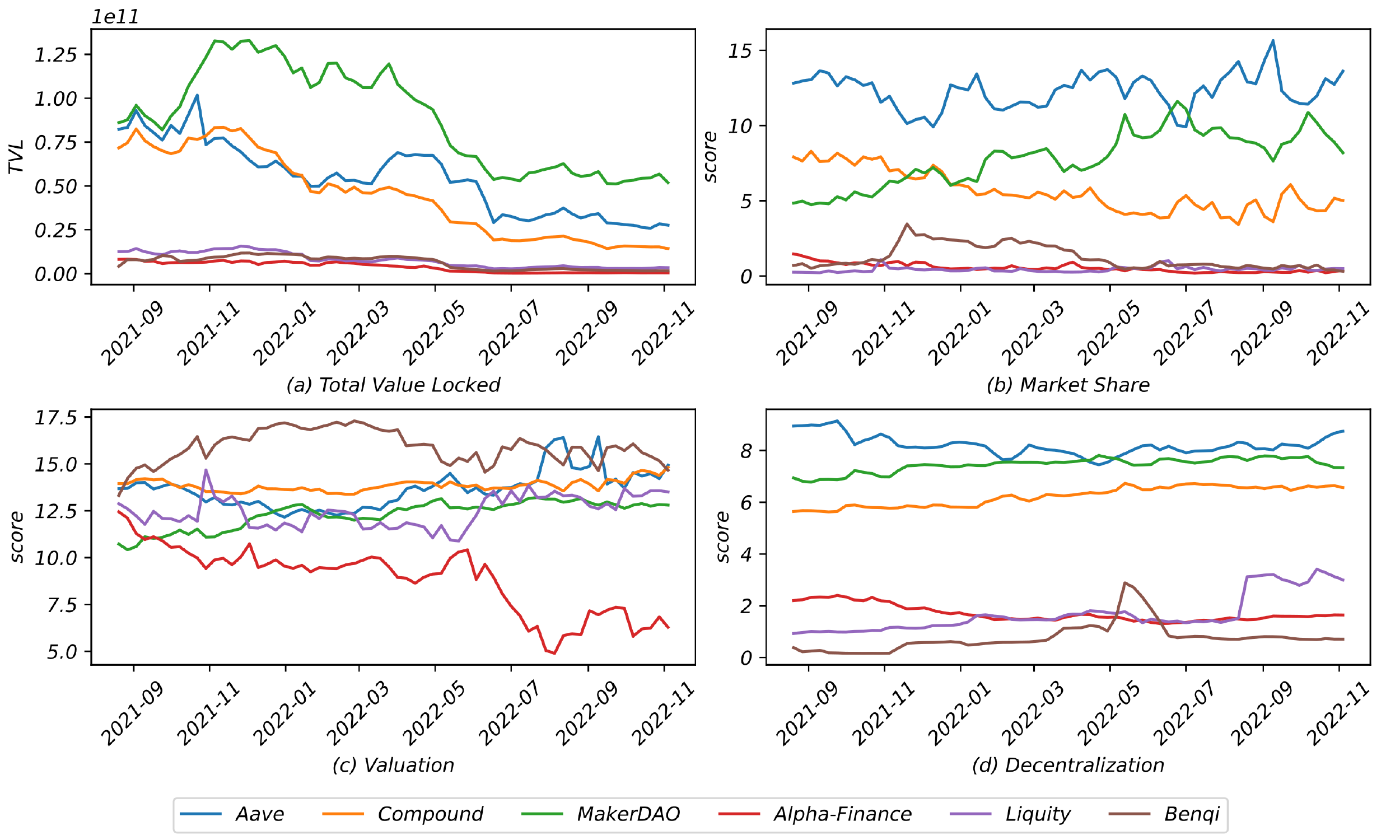}
	\caption{Comparison with the single total locked value ranking.}
	\label{fig:first}
\end{figure}

\subsection{Analysis of First-level Indicators}
\textit{1)  Comparison with single total value locked:}
The comparison between lending protocols evolution and first-level indicators is
shown in figure \ref{fig:first}. 
Compared with the TVL indicator, the first-level indicator takes into account market share and other related indicators of the protocol. Therefore, the first-level indicator can reflect more information about the protocol.
For example, the price of ETH has sharply declined from early may 2022, borrowers of lending protocol usually choose ETH as collateral to lend cryptocurrencies, therefore, the total locked value of lending protocols has also declined from May 2022. But from the curve of first-level indicators, we can discover more information. As shown in figure \ref{fig:first}(b), in addition to Aave and MakerDAO, although the total locked value of protocol has
declined, the market share of protocol has not changed drastically. However, the market share score of MakerDAO exceeded Aave. In general, it shows that the falling price of ETH affected the user's choice for different protocols. Similarly, in figure\ref{fig:first}(c) and figure\ref{fig:first}(d), We found some fluctuations in alpha-finance's valuation score and Benqi's decentralization score. Next, We will analyze the fluctions for each protocol in detail.

\begin{figure*}[htbp]
	\centering
	\includegraphics[width=18cm,trim=50 0 0 40]{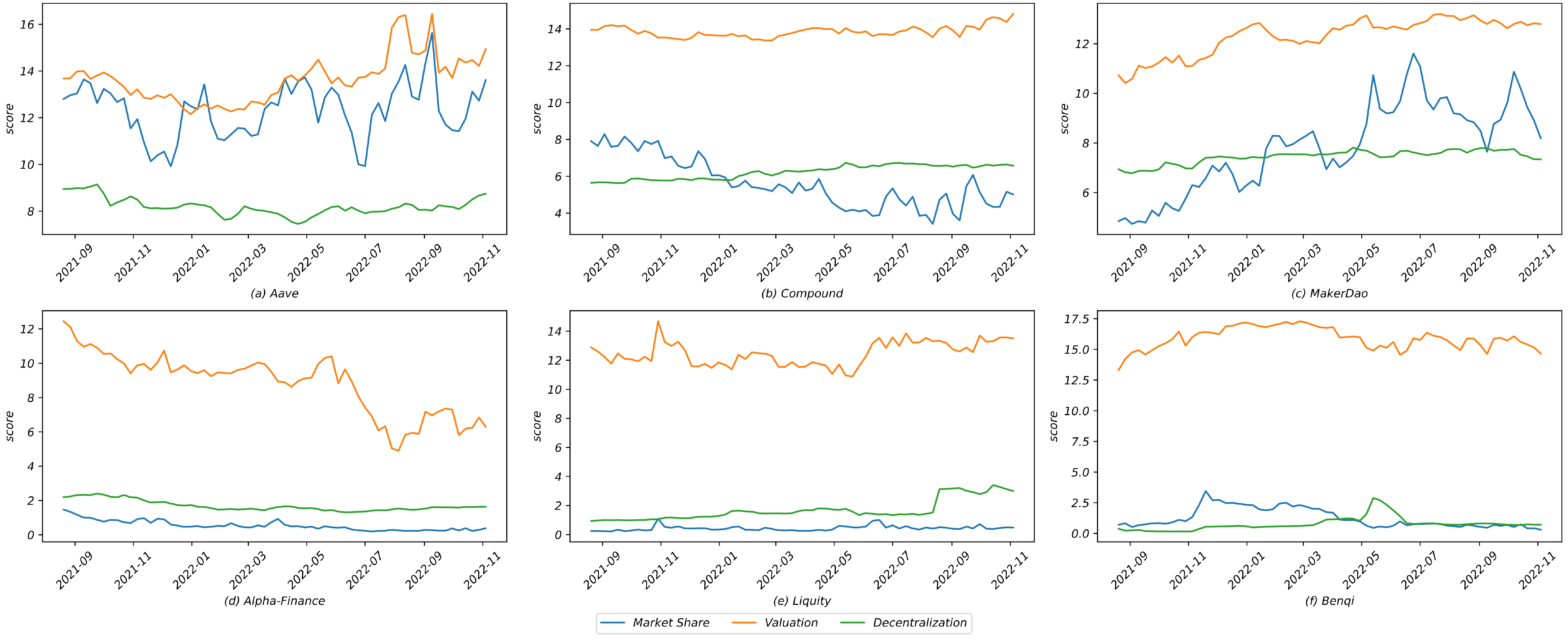}
	\caption{The evolution of the first-level indicators of single lending protocol.}
	\label{fig:protocol}
\end{figure*}

\textit{2)  The performance of a single lending protocol: } Figure \ref{fig:first} shows the fist-level indicators evolution of six lending protocols. Through the comparative analysis of multiple subgraphs, it is easy to see that the first-level indicators are uncorrelated. For example, in figure \ref{fig:protocol}(a) and \ref{fig:protocol}(c), it can be clearly seen that there is no correlation between Decentralization and Stability.

\begin{figure*}[htbp]
	\centering
	\includegraphics[width=17cm,trim=50 0 0 10]{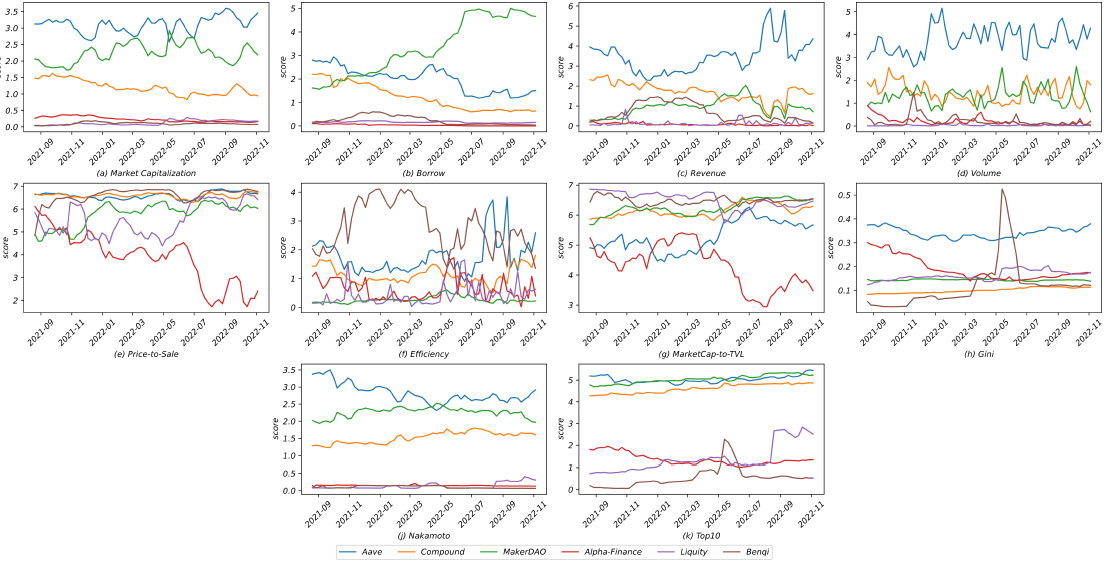}
	\caption{The evolution of the second-level indicators of single lending protocol.}
	\label{fig:firstIndex}
\end{figure*}

Other in-depth information can also be drawn from the analysis of figure \ref{fig:protocol}. For example, figure \ref{fig:rank}(b) shows that the rank of Aave protocol is in the first place most of the time, but there was some fluctuation in the total score around May 2022. The score of MakerDao exceeded Aave, and ranked first. Next, we will explain the reason for this fluctuation. From figure \ref{fig:protocol}, we can clearly see that the fluctuation of the Aave score mainly comes from the first level indicator of the market share around May. Then we further explain the fluctuation first-level indicators by analyzing the second-level indicators in figure \ref{fig:firstIndex}. In figure \ref{fig:firstIndex} (a) and figure \ref{fig:firstIndex} (b), both market capitalization score and borrow score of Aave have a sharp decline compared to other protocols. On the contrary, as shown in the figure \ref{fig:protocol}(c), the first-level indicator of market share of MakerDao was rising rapidly in the same time. Further more, it can be seen from the figure \ref{fig:firstIndex}(a) that the market capitalization score of MakerDAO far exceeded other protocols around May. 

From another perspective, because the price of ETH dropped sharply in early May, DeFi users tended to deposits cryptocurrencies as collateral in MakerDAO, and borrow stablecoin to avoid the risk of further depreciation of the collateral. In the same period, as shown in figure \ref{fig:firstIndex}(d), the token volume score of MakerDAO is growing rapidly. With the sale of AAVE tokens and the purchase of MKR tokens, the market capitalization score of MakerDAO exceeded the Aave. In addition, we can also analyze the impact of token transactions on the first-level indicator of decentralized. In figure \ref{fig:protocol}(a), the decentralization score of Aave showed an upward trend in May. By oberserving Decentralization-related secondary indicators in figure \ref{fig:firstIndex}, the top10 score and gini coefficient score of Aave was little changed, but the Nakamoto coefficient score has increasing slowly from May.

\subsection{Analysis of First-level Indicators}
Figure \ref{fig:firstIndex} shows the evolution of secondary indicators of lending protocols. Through comparative analysis, we can discover the trend in the development of the lending field. From the figure \ref{fig:firstIndex}(a) and \ref{fig:firstIndex}(b), no matter how the market capitalization of the MakerDAO and the price of ETH fluctuate, the proportion of borrow amount of MakerDAO has been increasing from the beginning of September 2021. Correspondingly, the incentives of Aave and Compound  are less attractive to DeFi users.

Moreover, we also discover some characteristics of decentralization. In figure \ref{fig:firstIndex}(h), the gini coefficient of the Benqi protocol has a violent fluctuation. The Benqi protocol has a dramatic fluctuation in the gini coefficient, which measures the overall distribution of tokens and generally does not fluctuate dramatically. Combined with Figure \ref{fig:firstIndex}(k), it can be found that indicator top10 and gini coefficient have a synchronous fluctuation, and the reason for the fluctuation of the decentralization may be that wallet addresses holding a large number of tokens trade off their tokens in exchange, making the tokens more evenly distributed.

\section{Conclusion}\label{sec:conclusion}

This paper established a four-layer evaluation model based
on AHP, and conducted a comprehensive evaluation of mainstream lending protocols. We collected and processed the data of the lending protocols that obtained from the Etherscan and Token Terminal. Based on the above data, we proposed three first-level indicators and ten second-level indicators covering all aspects of lending protocols. And we calculated the scores of six mainstream lending protocols from August 2021 to November 2022. From the visual analysis of calculation results, we found that the massive drop in the price of ETH increase the market share of stablecoin in the lending protocols and explained why it happened combined with the evaluation model. In addition, we also discovered indeepth reasons for fluctuations in the scores of lending protocols.

In our future work, we aim to quantify the risk of the lending protocols, add it to the evaluation model of the lending protocols, and then conduct a comprehensive analysis of the lending protocols to provide participants in the DeFi protocols with a correct and intuitive reference.


\begin{thebibliography}{0}

\bibitem{cefi}
Qin K, Zhou L, Afonin Y, et al, CeFi vs. DeFi--Comparing Centralized to Decentralized Finance, \emph{arXiv preprint}, 2021, arXiv:2106.08157.

\bibitem{santos}
S. Dos Santos, J. Singh, R. K. Thulasiram, S. Kamali, L. Sirico and L. Loud, A New Era of Blockchain-Powered Decentralized Finance (DeFi) - A Review, \emph{
2022 IEEE 46th Annual Computers, Software, and Applications Conference (COMPSAC)}, 2022: 1286-1292.

\bibitem{Salami}
Salami I, Challenges and approaches to regulating decentralized finance, \emph{American Journal of International Law}, 2021: 425-429.

\bibitem{DeFi-ning DeFi?Challenges & Pathway}
Amler H, Eckey L, Faust S, et al, Defi-ning defi: Challenges \& pathway, \emph{2021 3rd Conference on Blockchain Research \& Applications for Innovative Networks and Services (BRAINS)}, IEEE, 2021: 181-184.

\bibitem{SoK?Decentralized Finance (DeFi)}
Werner, Sam M., et al, Sok: Decentralized finance (defi), \emph{arXiv preprint}, arXiv:2101.08778, 2021

\bibitem{Busayatananphon}
C. Busayatananphon and E. Boonchieng, Financial Technology DeFi Protocol: A Review, \emph{2022 Joint International Conference on Digital Arts, Media and Technology with ECTI Northern Section Conference on Electrical, Electronics, Computer and Telecommunications Engineering (ECTI DAMT \& NCON)}, 2022: 267-272

\bibitem{Schar}
Sch\"{a}r F, Decentralized finance: On blockchain-and smart contract-based financial markets, \emph{FRB of St. Louis Review}, 2021.

\bibitem{llama}
DeFiLlama, Available: https://defillama.com/categories/

\bibitem{loanable funds}
Gudgeon L, Werner S, Perez D, et al, Defi protocols for loanable funds: Interest rates, liquidity and market efficiency, \emph{Proceedings of the 2nd ACM Conference on Advances in Financial Technologies}, 2020: 92-112.

\bibitem{Zhao}
Zhao Y, Kang X, Li T, et al, Toward trustworthy defi oracles: Past, present, and future, \emph{IEEE Access}, 2022, 10: 60914-60928.

\bibitem{Caldarelli}
Caldarelli G, Ellul J, The blockchain oracle problem in decentralized finance?a multivocal approach, \emph{Applied Sciences}, 2021, 11(16): 7572.

\bibitem{Liquidation}
Qin K, Zhou L, Gamito P, et al, An empirical study of defi liquidations: Incentives, risks, and instabilities, \emph{Proceedings of the 21st ACM Internet Measurement Conference}, 2021: 336-350.

\bibitem{Fenu}
G. Fenu, L. Marchesi, M. Marchesi and R. Tonelli, The ICO phenomenon and its relationships with ethereum smart contract environment, \emph{2018 International Workshop on Blockchain Oriented Software Engineering (IWBOSE)}, Campobasso, Italy, 2018: 26-32.

\bibitem{The decentralized crises}
Gudgeon, Lewis, et al, The decentralized financial crisis, \emph{2020 crypto valley conference on blockchain technology (CVCBT)}, IEEE, 2020.


\bibitem{d5}
Gochhayat S P, Shetty S, Mukkamala R, et al, Measuring decentrality in blockchain based systems, \emph{IEEE Access}, 2020: 178372-178390.


\bibitem{d3}
Nadler M, Sch\"{a}r F, Decentralized finance, centralized ownership? an iterative mapping process to measure protocol token distribution, \emph{arXiv preprint},2020, arXiv:2012.09306.

\bibitem{d4}
Lin, Q., Li, C., Zhao, X., \& Chen, X, Measuring decentralization in bitcoin and ethereum using multiple metrics and granularities, \emph{2021 IEEE 37th International Conference on Data Engineering Workshops (ICDEW)}, 2021: 80-87.


\bibitem{d1}
Zhang L, Ma X, Liu Y, SoK: Blockchain Decentralization, \emph{arXiv preprint},2022,arXiv:2205.04256.

\bibitem{Goodell}
Corbet S, Goodell J W, Gunay S, et al, Are DeFi tokens a separate asset class from conventional cryptocurrencies?, \emph{Annals of Operations Research}, 2023: 1-22.

\bibitem{Corbet}
Corbet S, Goodell J W, G\"{u}nay S, What drives DeFi prices? Investigating the effects of investor attention, \emph{Finance Research Letters}, 2022, 48: 102883.

\bibitem{Maouchi}
Maouchi Y, Charfeddine L, El Montasser G, Understanding digital bubbles amidst the COVID-19 pandemic: Evidence from DeFi and NFTs, \emph{Finance Research Letters}, 2022, 47: 102584.

\bibitem{Karim}
Karim S, Lucey B M, Naeem M A, et al, Examining the interrelatedness of NFTs, DeFi tokens and cryptocurrencies, \emph{Finance Research Letters}, 2022, 47: 102696.

\bibitem{Chousa}
Pi\~{n}eiro-Chousa J, L\'{o}pez-Cabarcos M \'{A}, Sevic A, et al, A preliminary assessment of the performance of DeFi cryptocurrencies in relation to other financial assets, volatility, and user-generated content, \emph{Technological Forecasting and Social Change}, 2022, 181: 121740.

\bibitem{topsis}
Tang H, Shi Y, Dong P, Public blockchain evaluation using entropy and TOPSIS, \emph{Expert Systems with Applications}, 2019, 117: 204-210.

\bibitem{Ozili}
Ozili P K, Decentralized finance research and developments around the World, \emph{Journal of Banking and Financial Technology}, 2022: 1-17.

\bibitem{Hartmann}
Hartmann J, Hasan O, Privacy considerations for a decentralized finance (DeFi) loans platform, \emph{Cluster Computing}, 2022: 1-15.

\bibitem{Etherscan}
 Etherscan, Avaliable: https://docs.etherscan.io/.
 
 \bibitem{tokenterminal}
 TokenTerminal, Avaliable: https://tokenterminal.com/

\end{thebibliography}
\end{document}